\documentclass[12pt,epsfig]{article}

\pdfoutput=1
\usepackage{sitender}
\usepackage[margin=2cm]{geometry}
\usepackage{comment}

\input epsf.sty
\renewcommand{\arraystretch}{1.5}

\title{2 Point tree Amplitudes in Pure Spinor }
\author{}

\begin{document}

	\baselineskip 24pt
	
	\begin{center}
		{\Large \bf  {Two-Point Superstring Tree Amplitudes Using the Pure Spinor Formalism}}
		
	\end{center}
	
	\vskip .3cm
	\medskip
	
	\vspace*{.10ex}
	
	\baselineskip=18pt

	\begin{center}
		
		{Sitender Pratap Kashyap$^{a,b}$}
		
	\end{center}
	
	\vspace*{2.0ex}

\begin{center} 		\it $^{a}$Institute of Physics\\
		Sachivalaya Marg,
		Bhubaneshwar 751005, India
\end{center}

	\centerline{ \it \small $^b$ Chennai Mathematical Institute, H1 SIPCOT IT Park,}
	\centerline{\it \small   Kelambakkam, Tamil Nadu, India 603103}
\vspace*{.5ex}
	\centerline{ E-mail:  sitender@cmi.ac.in}		
	\vspace*{5.0ex}

	\centerline{\bf Abstract} \bigskip
	We provide a prescription for computing two-point tree amplitudes in the pure spinor formalism that are finite and agree with the corresponding expression in the field theories. In \cite{Erbin:2019uiz,Seki:2019ycz}, {same} results were presented for bosonic strings and it was mentioned they can be generalized to superstrings. The pure spinor formalism is a successful super-Poincare covariant approach to quantization of superstrings \cite{Berkovits}.   Because the pure spinor formalism is equivalent to other superstring formalisms, we explicitly verify the above claim. {We introduce a {\it mostly BRST exact operator} in order to achieve this.} 
	\vfill
	
	
	\vfill \eject
	
	\baselineskip 18pt
	
	\tableofcontents
	
\section{Introduction}

The two point tree level bosonic string amplitudes in flat spacetime have been shown to be equal to the corresponding free particle expression in the quantum field theories \cite{Erbin:2019uiz,Seki:2019ycz}. We anticipate the same to be true in the superstrings as well.  In \cite{Erbin:2019uiz}, it was argued that their analysis will be carry over identically to the $NS-NS$ sector of superstrings. Further it was anticipated that spacetime supersymmetry will ensure similar story will repeat for other sectors of superstrings. It is desirable if this claim can be shown to be true in other formalism of superstrings as well. In this work we shall focus on the pure spinor formalism which keeps the Poincare and spacetime supersymmetry manifest \cite{Berkovits}. A naive application of the pure spinor amplitude prescription gives a vanishing two-point tree level amplitude. We shall see why this is the case and how to rectify the amplitude prescription. In what follows we briefly recall why the two point amplitudes in bosonic string were traditionally believed to vanish.   

In the bosonic string theory we have a diff $\times$ Weyl symmetry. This symmetry can be locally fixed by fixing the components of the worldsheet metric. The choice of metric however does not completely fix all the symmetries. These unfixed symmetries form the conformal killing group. On a sphere/disk this group is non-compact. For amplitudes involving three or more strings, this symmetry group is completely fixed by fixing the position of three of the vertex operators. In a two point amplitude, even after fixing the position of the two of the vertex operator, the residual conformal killing group still has infinite volume. This volume appears in the denominator of the corresponding path integral which naively implies that the path integral vanishes. Thus, these amplitudes were for a very long time assumed to vanish. This understanding however relies on the assumption that the numerator of the corresponding path integral in finite. In \cite{Erbin:2019uiz} it was noticed for the first time that the numerator is also $\inf$ and hence one has to make sense of an expression of the form $\f{\inf}{\inf}$. It was further shown that this expression is finite and gives rise to the expected two point amplitudes. It was also mentioned that same arguments can be applied to RNS superstrings.

 In \cite{Seki:2019ycz} these amplitudes were revisited in the operator formalism of bosonic strings. In this formalism unless three or more vertex operators are inserted, the amplitude vanishes because {of the non-saturation of ghost number} \cite{Polchinski}.  The two point amplitudes vanish on disk/sphere because the vertex operators supply two $c$ ghosts, while saturation of $c$ ghost zero modes require three $c$ ghosts. To get a finite result, the authors  introduce a novel vertex operator that statures the ghost zero modes in these amplitudes and provides the desired two point result.  They called it a {\it mostly BRST exact} operator. {We also follow this approach and introduce the operator 
	
	\be
	V_0(z)\equiv \f{1}{2\pi\a'}\int_{-\inf}^{\inf}dq \rb{\l\g^0\t}e^{iqX^0(z)} \label{v0}
	\ee
which is a mostly BRST exact operator and provides a non vanishing two point tree level amplitudes in the pure spinor formalism. The integrand of the above operator for $q\ne0$ can be written as $\sqb{Q,\star}$. This is why it is called {\it mostly} BRST exact (see equation \eqref{BRST-exact} for explicit form).  Similar operators have been used earlier but in a different context in \cite{Jusinskas:2015eza}.}

 Since $V_0$ explicitly uses space-time index, we must check for Lorentz invariance of the amplitude upon insertion of $V_0$ in the corresponding correlation function. It is important to ensure that insertion of $V_0$ does not alter the various symmetries of the amplitudes namely - super-Poincare and conformal invariance.  In the appendix \ref{consistency} we explicitly check that these symmetries are preserved\footnote{Upto BRST exact terms, which give vanishing contribution.}. Using this we shall freely use the consequences of these symmetries for reach various conclusions.

Rest of the paper is organized as follows. In section \ref{2pt_PS}  we first address why the two point amplitudes in the pure spinor formalism vanish on naive use of the standard prescription, then we introduce the mostly BRST operator and show that its use gives rise to the expected two point result. We end with a discussion in \ref{discuss} and defer some of details and computations to the appendices.

\section{Two point amplitudes in the pure spinor formalism} \label{2pt_PS}

In this section we begin by addressing the problem with two point amplitudes in the pure spinor formalism. We shall consider open strings for simplicity as the generalization to other string theories is straightforward - we shall make some comments on this in the discussion section \ref{discuss}. The amplitude prescription at tree level in the pure spinor formalism is given by \cite{Berkovits}
\be 
{\mc{A}_n=\int\prod_{j=4}^n d z_j\la V_1(z_1) V_2(z_2) V_3(z_3)U_j(z_j)\ra_{D^2}} \label{Npt}
\ee
where, $V$ are the unintegrated vertex operators and $U$ are the integrated vertex operators. We shall not present full details of the pure spinor formalism here, but give a quick review in \ref{review} containing the important ingredients required in this work - see \cite{Berkovits_ictp,Joost_thesis,Mafra_thesis,Oliver_thesis} for detailed reviews. We shall be concerned only with the unintegrated vertex operators which we take to be in the plane wave basis. They are given by 
\be 
V(z)=\hat{V} e^{ik.X}\equiv\l^\a O_\a e^{ik.X} \;,\quad\quad QV=0\;,\quad k^2=-\f{n}{\a'} \label{unintegrated}
\ee
where, $O_\a$ are conformal weight $n$, composite operators constructed out of the basic world-sheet fields $\Pi^m,d_\a,\t^a,N^{mn},J$. Also,  $n$ stands for the $n$-th excited level of the string and $Q$ denotes the BRST charge. The $e^{ik.X}$ cancels the conformal weight of $O_\a$ so that $V$ has zero conformal dimension.

 We should note that there is so far no derivation of amplitude prescription \eqref{Npt} in the absence of the underlying gauge theory for pure spinor formalism whose gauge fixing gives \eqref{Npt}\footnote{Recently a gauge theory behind the pure spinor formalism was proposed in \cite{Jusinskas:2019vmd}. Perhaps one can arrive at the amplitude prescription using this. This has not been done so far.}. A justification for this amplitude prescription relies on the fact that the pure spinor formalism in its non-minimal version is $N=2$ topological strings \cite{Berkovits_topological} whose amplitude prescription is same as that of the bosonic strings\footnote{This amplitude prescription was derived in \cite{Hoogeveen:2007tu} by coupling the standard pure spinor formalism to topological gravity and performing a BRST quantization.}. 

All the non-trivial amplitudes in the pure spinor formalism can be brought to a form where there are three $\l$ and five $\t$ zero modes in the corresponding correlator. We choose to normalize all the amplitudes with respect to the following correlator\footnote{Normalizing this correlation function is sufficient since there is only one scalar present in tensor product of three $\lambda$ and five $\theta$.}\footnote{There is an alternative zero mode normalization for $\l$ and $\t$ given by $\la {\bf{1}}\ra_0=1$ \cite{Berkovits:2016xnb}. We shall not be working with this. See \cite{Bischof:2020tnf}) for an application of this prescription to compute one point closed string amplitudes on a disk. }
\be 
\la\rb{\l \gamma^m\theta} \rb{\l \gamma^m\theta} \rb{\l \gamma^p\theta}\rb{\theta \gamma_{mnp}\theta}\ra =1
\ee
For $n=2$, a naive application of \eqref{Npt} gives  
\be 
\mc{A}_2=\la V_1(z_1) V_2(z_2)\ra_{D^2} \propto \left\la \rb{\l^\a O^1_\a}(z_1) \; \rb{\l^\b O^2_\b}(z_2)\right\ra_{D^2}\label{2pt}
\ee
Since the above correlator has only two $\l$, it vanishes identically, implying that the above prescription gives a trivial two point amplitude. We must find out the correlator that gives rise to the correct two point scattering amplitude. This correlator must involve an extra {\it vertex operator} to be non-vanishing and at the same time gets rid of $\d(k_1^0-k_2^0)$. This suggests that the extra piece that we need must have one $\l^\a$. The operator \eqref{v0} fulfills these requirements. 

Let us begin by calculating the following amplitude
\be
A\equiv\la V_0(z)V_1(z_1) V_2(z_2)\ra=\f{1}{2\pi\a'}\int_{-\inf}^{\inf}dq\left\la\left[\rb{\l\g^0\t}(z)e^{iqX^0(z)}\right]\;V_1(z_1)V_2(z_2)\right\ra \label{2amp_prescription}
\ee
where, $V_0$ is the operator introduced in \eqref{v0} which we fix at $z$, while $V_1$ and $V_2$ are the unintegrated vertex operators (fixed at $z_1$ and $z_2$ respectively). 

In order to calculate this, we split the integral into three parts as follows
\be
A&=&\f{1}{2\pi\a'}\biggl(\int_{-\inf}^{-\e}dq\left\la\left[\rb{\l\g^0\t}(z)e^{iqX^0(z)}\right]\;V_1(z_1)V_2(z_2)\right\ra +\int_{-\e}^{\e}dq\left\la\left[\rb{\l\g^0\t}(z)e^{iqX^0(z)}\right]\;V_1(z_1)V_2(z_2)\right\ra \non\\
&&+\int_{\e}^{\infty}dq\left\la\left[\rb{\l\g^0\t}(z)e^{iqX^0(z)}\right]\;V_1(z_1)V_2(z_2)\right\ra\biggl) \non
\ee
The first and the third terms of the above equation vanish by replacing the operator in the square bracket with the BRST expression given in \eqref{BRST-exact}. This can be shown by unwrapping the contour around $z$, and circling $z_1$, $z_2$ and a contour containing $z,z_1,z_2$. The contour integrals around $z_1$ and $z_2$ vanish due to $QV_1=0$ and $QV_2=0$, while the remaining one can be deformed to include point at infinity which vanishes as well since there are no non-trivial operators there. Thus, we are left with 
\be
A=\f{1}{2\pi\a'}\int_{-\e}^{\e}dq\left\la\left[\rb{\l\g^0\t}(z)e^{iqX^0(z)}\right]\;V_1(z_1)V_2(z_2)\right\ra
\ee

On substituting the form of $V_i$ given in \eqref{unintegrated}, we can factor the amplitude as 
\be
A\equiv\f{1}{2\pi\a'}\int_{-\inf}^{\inf}dq\left\la\left\la\rb{\l\g^0\t}(z)\;\hat V_1(z_1)\hat V_2(z_2)\right\ra\right\ra \left\la e^{iqX^0}(z) e^{ik_1.X} (z_1) e^{-ik_2.X}(z_2)\right\ra \label{amp}
\ee
{where, we use the notation $\left\la\left\la\cdots \right\ra\right\ra$ as a shorthand to denote that the necessary OPE have been taken among the operators inside the bracket. We have taken the momentum $k_1$ to be incoming and $k_2$ to be outgoing.} The Koba-Nielsen factor of the above equation reduces to
\be
&&\hspace{-0.8in}\left\la e^{iqX^0}(z) e^{ik_1.X} (z_1) e^{-ik_2.X}(z_2)\right\ra\non\\
&=&iC_{D_2}^X\rb{2\pi}^{10}\d\rb{q+k_1^0-k_2^0}\;\d^{(9)}\big(\vec{k}_1-\vec{k}_2\big)|z-z_1|^{2\a'qk^0_1}|z-z_2|^{-2\a'qk^0_2}|z_1-z_2|^{-2\a'k_1.k_2} 
\ee  
On substituting the above result in \eqref{amp}, we find
\be
A&=&\f{i}{\a'}C_{D_2}^X  \int_{-\inf}^{\inf}dq\d\rb{q+k_1^0-k_2^0} (2\pi)^9\;\d^{(9)}\big(\vec{k}_1-\vec{k}_2\big)\left\la\left\la\rb{\l\g^0\t}(z)\;\hat V_1(z_1)\hat V_2(z_2)\right\ra\right\ra\non\\
&&\times  |z-z_1|^{2\a'qk^0_1}|z-z_2|^{-2\a'qk^0_2}|z_1-z_2|^{-2\a'k_1.k_2}
\ee
We are interested in the on-shell amplitudes for which $k^0=\sqrt{|\vec{k}|^2+m^2}$, for a particle carrying momentum $\vec{k}$ and mass $m$. The space Dirac-delta function in the above sets $\vec{k}_1=\vec{k}_2$. Further, let us assume that $m_2-m_1=\d$, where $m_1$ and $m_2$ are the masses of the string 1 and 2. If $m_1=m$ and $m_2=m+\d$, we find that on-shell 
$$k_2^0-k_1^0=\sqrt{|\vec{k}|^2+m^2}\sqb{\sqrt{1+\f{\d^2+2\d m}{|\vec{k}|^2+m^2}}-1}$$ However, since $-\e <q <\e$ , for getting a support from the energy Dirac-delta we must have 
$$-\e < \sqrt{|\vec{k}|^2+m^2}\sqb{\sqrt{1+\f{\d^2+2\d m}{|\vec{k}|^2+m^2}}-1} <\e .$$ Since, $\e\rightarrow 0$, we must have $\d\rightarrow 0$ for energy Dirac-delta to provide a non-trivial contribution. This means unless the masses of the strings are same, the amplitude vanishes on-shell. Thus, we find that 
\be
A=\f{i}{\a'}C_{D_2}^X  (2\pi)^9\;\d^{(9)}\big(\vec{k}_1-\vec{k}_2\big)\left\la\left\la\rb{\l\g^0\t}(z)\hat V_1(z_1)\hat V_2(z_2)\right\ra\right\ra \;|z_1-z_2|^{2\a'm_1^2} \;\d_{m_1,m_2} \label{amp1}
\ee
where, we used $k_1.k_2=-k_1^0 k_2^0+\vec{k}_1.\vec{k}_2=-(k_1^0)^2+|\vec{k}_1|^2=-m_1^2$.   
We see that there is a factor of $|z_{12}|^{-2\a m_1^2}$. This factor cancels with a similar factor coming from pure spinor superspace, namely $\left\la\left\la\rb{\l\g^0\t}\hat V_1(z_1)\hat V_2(z_2)\right\ra\right\ra$. Recall that at the $n^{th}$ level of open superstring we have 
$\rb{\text{mass}}^2=\f{n}{\a'}$ and also that the conformal dimensions of $\hat V_{i}$ for $i=1,2$  are $n$. Furthermore $(\l\g^0\t)$ has conformal weight zero. Thus, upon using the standard result for a 3-point function in a CFT, we find  
\be
\left\la\left\la\rb{\l\g^0\t}(z)\;\hat V_1(z_1)\hat V_2(z_2)\right\ra\right\ra \propto |z_{12}|^{-2n}
=|z_{12}|^{-2\a' m_1^2}\ee
which, cancels the coordinate dependence coming from the  Koba-Nielsen factor. Thus, the amplitude is coordinate invariant.

Having fixed the coordinate dependence, let us further elaborate on the dependence of $\la \la \cdots\ra \ra$ on the kinematic data i.e. the polarizations and momenta. We argue that it must be of the form 
\be 
\left\la\left\la\rb{\l\g^0\t}(z)\;\hat V_1(z_1)\hat V_2(z_2)\right\ra\right\ra \propto f^{0}(\e_1,\e_2;k) 
\ee
where  $\e_i$ are the polarizations  and $k$ is momentum of the state represented by vertex operators $V_1$ and $V_2$. Since our theory is supersymmetric, giving the argument for purely bosonic states is sufficient. The bosonic states will have their polarizations given by Lorentz vector indices. The polarizations satisfy $k_{m}\e_{i}^{m\cdots}=0$. Now, suppose that the $0$ index is supplied by $\e_1$. Then for non-zero answer, we can contract rest of the indices of $\e_1$ with only $\e_2$. But, this leaves a free index on $\e_2$ which must be contracted by $k$, giving a vanishing contribution. Thus, there is a uniqure choice - the polarization tensors contract among themselves and $0$ index is supplied by $k^0$.  In appendix \ref{2pt_calc} we explicitly verify this for all states at massless level.
Thus,  we find\footnote{There are other factors containing contribution of non-zero modes of various worldsheet fields, normalization of polarizations and pure spinor superspace computations. These are all non-zero.}  
\be 
\left\la\left\la\rb{\l\g^0\t}\hat V_1(z_1)\hat V_2(z_2)\right\ra\right\ra \propto k^0  \;\d_{jj'}  \label{k0_arg}
\ee
where, we have used $j$ and $j'$ in $\d_{jj'}$ to distinguish between states with degenerate masses like gluon and gluino. 
Hence, the final result  
\be
A\propto  (2\pi)^9\;\d^{(9)}\big(\vec{k}_1-\vec{k}_2\big) k^0\; \d_{m_1,m_2} \;\d_{jj'} \label{amp2}
\ee
reproduces the expected two point amplitude in a field theory in $D$ dimensions given by 
\be 
\mc{A}_2=2k^0\rb{2\pi}^{D-1}\d^{D-1}\big(\vec{k}_1-\vec{k}_2\big)\;, \qquad k^0\equiv\sqrt{m^2+\vec{k}^2}          \label{field_2pt}
\ee 
upto a proportionality constant which can be fixed by making use of a unitarity requirement
\be 
A_2(k_1,k_2)=\int \f{d^{D-1} k}{(2\pi)^{D-1}} A_2(k_1,k) A_2(k,k_2)
\ee In appendix \ref{2pt_calc} we explicitly verify the above result for some amplitudes. Thus, the two point amplitudes in the pure spinor formalism using operator $V_0$ behave as is expected.

\section{Discussion} \label{discuss}
{The non-vanishing of two point amplitudes is required for various consistencies, in particular at tree level in string theories - see \cite{Harold_SFT} for a discussion. }
We showed that by making use of an additional set of mostly BRST-exact states we can get non-zero two point tree amplitudes in the pure spinor formalism in open strings. 
Generalization to the closed strings is straight-forward by adding a right moving sector. Generalization to Heterotic strings follows by making use the pure spinor prescription of this work and for the bosonic sector by using the analysis of \cite{Erbin:2019uiz,Seki:2019ycz}.

To conclude we have identified the correlation functions that give rise to correct two point amplitude in the pure spinor formalism. We do not however know from a fundamental point of view why the additional vertex operator is of this form\footnote{Perhaps these are related to the zero momentum states as pointed out by Renann Lipinski Jusinskas. We note that the additional operators in \cite{Erbin:2019uiz,Seki:2019ycz} too resemble zero momentum states. The understanding of their precise role we leave for a future work.}. It is important to explore for a fundamental origin of the mostly BRST exact operator we used in this work, perhaps by making use of the gauge invariant action presented in \cite{Jusinskas:2019vmd} (see also \cite{Berkovits_BRSTaction,Berkovits:2015yra} which gave important insights that lead to \cite{Jusinskas:2019vmd}) . This investigation we leave for future work.  

\bigskip
\bigskip
\noindent{\bf Acknowledgments:}  
I wish to thank Biswajit Das for some discussions and Ashoke Sen and Mritunjay Verma for providing some useful comments on the draft. I am indebted to Renann Lipinski Jusinskas for many insightful discussions at the initial stages this work. I am thankful to Institute of Physics, Bhubaneshwar for generously providing a three month extension beyond the usual term of my post-doctoral tenure, during the pandemic due to CoVid-19.

\appendix
	\section{Review of the Pure Spinor Formalism} \label{review}

In this appendix, we very briefly review the pure spinor formalism in a way that makes the discussion in the main text coherent. The world sheet action in a conformal gauge for strings in a flat $10D$ spacetime takes the form
	\be
	S=\f{1}{\pi \alpha'}\int d^2z \left(\f{1}{2}\p X^m\bar\p X_m+p_\alpha\bar\p\theta^\alpha-w_\alpha\bar\p\lambda^\alpha\right)
	\ee
	where, $m\in\{0,1,\cdots,9\}$ and $\alpha \in \{1,\cdots,16\}$. 
 $X^{m}$ are the spacetime coordinates, $\theta^\alpha, w_\beta$ are anti-commuting Majorana-Weyl spinors while $\l^\a$ are commuting Weyl Spinors. $\{X^m,\theta^\alpha,\l^\a\}$ are scalars on the worldsheet, while $p_\alpha, w_\alpha$, the conjugate momenta fields of $\t^\a$ and $w_\a$ respectively, carry weight 1. Further, $\lambda^\alpha$ satisfy the {\it pure spinor}
	\be
	\lambda^\alpha\gamma^{m}_{\alpha\beta}\lambda^\beta=0\label{psconstr}  \qquad \forall \; m
	\ee
where, $\g^{m}_{\a\b}$ are symmetric $16\times 16$ Gamma matices in 10 dimensional spacetime. To keep the supersymmetry manifest, instead of working with $p_\alpha$ and $\p X^m$, we work with the supersymmetric combinations
\be
d_\alpha&=&p_\alpha-\f{1}{2}\gamma^m_{\;\;\alpha\beta}\theta^\beta\partial X_m-\f{1}{8}\gamma^m_{\alpha\beta}\gamma_{m\sigma\delta}\theta^\beta\theta^{\sigma}\partial\theta^\delta
\non\\[.4cm]\Pi^m&=&\partial X^m+\f{1}{2}\gamma^m_{\alpha\beta}\theta^\alpha\partial\theta^\beta\label{Pim}
\ee
The BRST operator is {\it postulated} to be
	\be
	Q=\oint dz\ \lambda^\alpha(z)\; d_\alpha (z)
	\ee
Due to the pure spinor constraint, $w_\alpha$ are defined upto a gauge transformation 
\be 
w_\a = \ul_{m}(\g^m\l)_\a
\ee 
with $\ul_{m}$ playing the role of gauge parameters. To take care of this gauge symmetry,  we always work with the following gauge invariant combinations
	\be
	N_{mn}=\f{1}{2} w_\alpha(\gamma_{mn})^\alpha_{\;\beta}\lambda^\beta\quad,\quad J=w_\alpha \lambda^\alpha\quad,\quad T= w_\alpha \p\lambda^\alpha\non
	\ee
$J$ is the ghost-number current and provides ghost number 1 to $\lambda^\a$.

The physical states lie in the BRST cohomology with ghost number 1.  The vertex operators are constructed out of $\{\Pi^m, d_\a, \t^\a, N^{mn}, J,\l^\alpha\}$.  The only non-trivia OPE we shall require is given by 
\be
d_\alpha(z)V(w)=\f{\alpha'}{2(z-w)}D_\alpha V(w)+\cdots 
\ee
where, $V$ denotes an arbitrary superfield while $D_\alpha$ is the supercovariant derivative given by
\be D_\alpha\equiv\p_\alpha+\gamma^m_{\alpha\beta}\theta^\beta\p_m
\quad\implies\qquad
\lbrace D_\alpha, D_\beta\rbrace= 2 (\gamma^m)_{\alpha \beta} \partial_m\label{cov_der}
\ee
where, $\p_{m}=\f{\p}{\p X^m}$ and $\p_{\a}=\f{\p}{\p \t^\a}$. The above describes what is called the {\it minimal} version of the pure spinor formalism and this will be sufficient for this work. 

As a matter of notation, we shall be implicit about normal ordering where a composite operators are defined via
\be 
:A(z)B(z): \equiv \frac{1}{2 \pi i} \oint_z  \frac{dw}{w-z} A(w)B(z)  \label{normal_order}
\ee
where, $A$ and $B$ are arbitrary operators.

\section{$V_0$ as a mostly BRST exact operator } \label{VisBRST-exact}
In this appendix we show that the $V_0$ introduced in the main body is a {\it mostly BRST-exact} operator. Let us begin by noticing
\be 
\sqb{Q,e^{iqX^0(z)}}\equiv \oint_{z}dw (\l^\a d_\a)(w) e^{iqX^0(z)}=\f{\a'}{2}\oint_zdw \l^\a(w)\sqb{ \f{D_\a e^{iqX^0(z)}}{w-z}+\cdots}
\ee
where, we used the standard OPE $d_\a(z) V(w)\simeq \f{\a'}{2}\f{D_\a V}{z-w}$. On recalling that $D_\a=\p_\a+\rb{\g^m\t}_\a \p_m$, we find that 
\be
\sqb{Q,e^{iqX^0(z)}}=\f{iq\a'}{2}\rb{\l\g^0\t} e^{iqX^0}
\ee
Hence, for $q\ne 0$ we have 
\be 
\rb{\l\g^0\t}e^{iqX^0}=-\f{1}{q}\sqb{Q,\rb{\f{2i}{\a'}e^{iqX^0(z)}} }  \label{BRST-exact}
\ee
showing that the integrand of $V_0$ is BRST-exact for $q\ne 0$ and thus $V_0$ is mostly BRST exact. 

\section{Some explicit examples} \label{2pt_calc}
In this appendix we explicitly calculate the two point massless open superstring amplitudes  (on a disk) with the new prescription given equation \eqref{2amp_prescription} in this paper. The goal is to substantiate the claim in \eqref{k0_arg} by providing some explicit examples. For this we essentially need to compute the $\left\la\left\la \rb{\l\g^0\t}\hat{V}_1 \hat{V}_2\right\ra\right\ra$ where $\hat{V}_i=\l^\a A_{i\a}$ and show it is proportional to $k^0$. For the massless case the task is trivial as everything inside the bracket is conformal weight zero and hence there are no non-trivial OPEs. Consequently we can directly use the pure spinor-superspace method to perform the computation. The relevant theta expansion is given by (we follow the notation and conventions used in \cite{Equivalence})
\be 
A_\a=a_m\rb{\g^m}_\a -\f{2}{3}\rb{\g^m\t}_\a \rb{\t\g_m\c}+\cdots
\ee 
where, we have not shown the higher order $\t$ terms as they will not be required. Also, $a_m$ represents the gluon field and $\c^\a$ the gluino field.  Consequently we find
\be 
\left\la\left\la \rb{\l\g^0\t}\hat{V}_1 \hat{V}_2\right\ra\right\ra=\left\la \rb{\l\g^0\t}\rb{\l^\a A_{1\a}}\rb{ \l^\a A_{2\a}}\right\ra =\f{i}{180}k^0 \rb{a_r^m a_{r'm}}+\f{1}{360}\rb{\c_s\g^0\c_{s'}}
\ee
where, $r,r'$ and $s,s'$ denote polarizations and helicities of gluons and gluinos respectively. 
We made use of the following pure spinor superspace identities \cite{Berkovits:2006bk}, for performing the gluon calculation
\be 
\la\rb{\l\t\g^m\r} \rb{\l\t\g^n\r} \rb{\l\t\g^p\r}\rb{\l\t\g_{s t u}\r}\ra=\f{1}{120}\delta^{m n p}_{s t u}
\ee
and 
\be
&&\hspace*{-.35in}\langle(\lambda\gamma^{u}\theta)(\theta\gamma_{fgh}\theta)(\theta\gamma_{jkl}\theta)(\lambda\gamma_{mnpqr}\lambda)\rangle\non\\[.3cm]
&=&-\f{4}{35}\Bigl[\delta^{[m}_{[j}\delta^{n}_{k}\delta^{p}_{l]}\delta^{q}_{[f}\delta^{r]}_{g}\delta^{u}_{h]}+\delta^{[m}_{[f}\delta^{n}_{g}\delta^{p}_{h]}\delta^{q}_{[j}\delta^{r]}_{k}\delta^{u}_{l]}-\f{1}{2}\delta^{[m}_{[j}\delta^{n}_{k}\eta_{l][f}\delta^{p}_{g}\delta^{q}_{h]}\eta^{r]u}-\f{1}{2}\delta^{[m}_{[f}\delta^{n}_{g}\eta_{h][j}\delta^{p}_{k}\delta^{q}_{l]}\eta^{r]u}\Bigl]\non\\
&&-\f{1}{1050}\epsilon^{mnpqr}_{\;\;\;\;\;\;\;\;\;\;\;\;abcde}\left[\delta^{[a}_{[j}\delta^{b}_{k}\delta^{c}_{l]}\delta^{d}_{[f}\delta^{e]}_{g}\delta^{u}_{h]}+\delta^{[a}_{[f}\delta^{b}_{g}\delta^{c}_{h]}\delta^{d}_{[j}\delta^{e]}_{k}\delta^{u}_{l]}-\f{1}{2}\delta^{[a}_{[j}\delta^{b}_{k}\eta_{l][f}\delta^{c}_{g}\delta^{d}_{h]}\eta^{e]u}-\f{1}{2}\delta^{[a}_{[f}\delta^{b}_{g}\eta_{h][j}\delta^{c}_{k}\delta^{d}_{l]}\eta^{e]u}\right]\non\\
\ee
for gluino calculation\footnote{We acknowledge the use of \cite{Peeters1,Gran} for performing the calculations.}. We further note that the polarizations are normalized as (see for example \cite{CaronHuot:2010rj})
\be 
a_r^m a_{r'm}=\d_{rr'}\;,\quad \rb{\c_s\g^0\c_{s'}}=k^0\d_{ss'}
\ee
Thus, we find that 
\be 
\left\la\left\la \rb{\l\g^0\t}\hat{V}_{1j} \hat{V}_{2j'}\right\ra\right\ra\propto k^0 \d_{j j'}\d_{s s'}
\ee
where, $j,j'$ stand for the particle species and $s,s'$ denote the corresponding polarizations. We note that we get the same relative factor in gluon and gluino amplitude as in \cite{Equivalence} and that the gluon-gluino amplitudes vanish automatically as we expect them to. For the massive case the computations can be repeated using \cite{Equivalence,Unintegrated}, though we expect them to be involved. 

\section{Some consistency checks for $V_0$ insertion}  \label{consistency}
In this appendix we verify that the insertion of $V_0$ does not spoil the super-Poincare and conformal invariance. First note that the expression of $V_0$ is not Lorentz covariant as it isolates $0th$ spacetime component. Superstring theory in the flat background is super-Poincare invariant. So, let us first see if the expression that we have arrived has this symmetry. Notice that 
\be
\d\la V_0 V_1 V_2\ra=\la \d V_0 V_1 V_2\ra+\underbrace{\la V_0 \d V_1 V_2\ra+\la V_0 V_1 \d V_2\ra}_{\text{automatically invariant}}=\la \d V_0 V_1 V_2\ra
\ee
where, $\d$ denotes the change after application of some symmetry transformation. The vertex operators $V_1$ and $V_2$ are by definition invariant under $\d$ and hence we only need to evaluate $\d V_0$.
In order to facilitate the discussion, let us note that we can write $V_0$ as 
\be 
V_0=\int_{-\inf}^{\inf} \f{dq}{i\pi {\a'}^2}\f{1}{q}\sqb{Q,e^{iqX^0}}
\ee
Now, we can write the variations as (on noticing that $\d Q=0$ ) 
\be
\d V_0=\int_{-\inf}^{\inf} \f{dq}{i\pi {\a'}^2}\f{1}{q}\sqb{Q,\d e^{iqX^0}}
\ee 
Let us now consider all the transformations one by one. To distinguish one transformation from other, we shall provide a subscript on $\d$. Under translations 
\be 
X^m\rightarrow X^m+a^m \implies \quad \d_{a} X^m=a^m \implies \d_a e^{iqX^0}=iqa^0e^{iqX^0}
\ee
Thus, we see that 
\be 
\d_{a} V_0 =\int_{\inf}^{\inf} \f{dq}{i\pi {\a'}^2}\f{1}{q} \,\;\sqb{Q, iq a^0e^{iqX^0}}=\sqb{Q,  \int_{\inf}^{\inf} \f{dq}{\pi {\a'}^2}\; a^0e^{iqX^0}}
\ee
On making use of $QV_1=0=QV_2$, we can easily see that the two point amplitude is translationally invariant. Similarly, under Lorentz transformations 
\be 
X^m\rightarrow X^m+\ul^{m}_{\;\;n}X^n \quad \implies\quad\d_{\ul} V_{0}=\sqb{Q, \int_{\inf}^{\inf} \f{dq}{\pi {\a'}^2} \ul^0_{m}X^me^{iqX^0}}
\ee
and under supersymmetry transformation 
\be 
X^m\rightarrow X^m+\rb{\eta\g^m\t}\quad \implies \quad \d_{\eta} V_{0}=\sqb{Q, \int_{\inf}^{\inf} \f{dq}{\pi {\a'}^2} \rb{\eta\g^m\t}e^{iqX^0}}
\ee
Thus, we see that the amplitude is super-Poincare invariant. Finally under conformal transformations $z\rightarrow z+\e(z)z\implies \d X^m=\e(z)\p X^m(z)$, we have
\be 
\d_{\e} V_{0}=\sqb{Q, \int_{\inf}^{\inf} \f{dq}{\pi {\a'}^2}  \e(z)e^{iqX^0}}
\ee 
showing that the two point amplitude is conformally invariant.
\bibliographystyle{JHEP}
\bibliography{reference}
\end{document}